\documentclass[12pt,a4paper]{article}
\usepackage{amssymb,amsmath}
\usepackage{ifxetex}
\usepackage{fixltx2e} 
\ifxetex
  \usepackage{fontspec,xltxtra,xunicode}
  \usepackage{mathpazo}
  \defaultfontfeatures{Mapping=tex-text,Scale=MatchLowercase}
  
  \usepackage[a4paper,width=36em,height=72em]{geometry}
\else
  \usepackage[utf8]{inputenc}
  \usepackage{mathpazo}
  \usepackage[letterpaper,width=30em,height=56em]{geometry}
\fi
\usepackage{fancyvrb}
\usepackage{url}

\usepackage{graphicx}
\let\Oldincludegraphics\includegraphics
\renewcommand{\includegraphics}[1]{\makebox[0pt]{\Oldincludegraphics{#1}}}

\usepackage[font=small,format=plain,labelfont=bf,width=190mm]{caption}

\ifxetex
  \usepackage[setpagesize=false, 
              unicode=true,
              xetex,
              bookmarks=true,
              pdfpagemode=UseNone,
              pdfstartview=FitBH,
              pdfauthor={Carsten Allefeld\^{}1,2*\^{}; John-Dylan Haynes\^{}1--6**\^{}},
              pdftitle={Searchlight-based  multi-voxel pattern analysis of fMRI  by cross-validated MANOVA}]{hyperref}
\else
  \usepackage[unicode=true,
              bookmarks=true,
              pdfpagemode=UseNone,
              pdfstartview=FitBH,
              pdfauthor={Carsten Allefeld\^{}1,2*\^{}; John-Dylan Haynes\^{}1--6**\^{}},
              pdftitle={Searchlight-based  multi-voxel pattern analysis of fMRI  by cross-validated MANOVA}]{hyperref}
\fi
\hypersetup{breaklinks=true, pdfborder={0 0 0}}
\usepackage{sectsty}
\allsectionsfont{\sffamily\raggedright}
\subparagraphfont{\sffamily\mdseries}

\DeclareRobustCommand{\[}{\begin{equation}}
\DeclareRobustCommand{\]}{\end{equation}}

\renewcommand{\eqref}[1]{(Eq.\ \ref{#1})}

\newcommand{\mathup}[1]{\mathrm{#1}}

\DeclareMathOperator{\var}{var}         
\DeclareMathOperator{\rk}{rank}         
\DeclareMathOperator{\tr}{trace}        
\DeclareMathOperator{\Hnull}{H_0}       
\let\dist\sim                           
\newcommand{\Dist}[1]{\mathcal{#1}}     

\newcommand{\A}{\mathup A}
\newcommand{\B}{\mathup B}
\newcommand{\E}{\mathup E}
\newcommand{\I}{\mathup I}

\title{\sffamily\bfseries Searchlight-based\\ multi-voxel pattern analysis of fMRI\\ by
cross-validated MANOVA}  
\author{\sffamily\mdseries Carsten Allefeld\^{}1,2*\^{}; John-Dylan Haynes\^{}1--6**\^{}}
\date{}     

\begin{document}
\maketitle

\begin{scriptsize}
Affiliations:\\
1. Bernstein Center for Computational Neuroscience, Charité – Universitätsmedizin Berlin, Germany\\
2. Berlin Center of Advanced Neuroimaging, Charité – Universitätsmedizin Berlin, Germany\\
3. Department of Neurology, Charité – Universitätsmedizin Berlin, Germany\\
4. Berlin School of Mind and Brain, Humboldt-Universität zu Berlin, Germany\\
5. Excellence Cluster NeuroCure, Charité – Universitätsmedizin Berlin, Germany\\
6. Department of Psychology, Humboldt-Universität zu Berlin, Germany

Address for all affiliations:\\
Charité-Campus Mitte\\
Philippstr. 13, Haus 6\\
10115 Berlin, Germany

* E-mail: carsten.allefeld@bccn-berlin.de\\
** E-mail: haynes@bccn-berlin.de

Corresponding author: Carsten Allefeld, Tel. +49 30 2093 6766
\end{scriptsize}

\begin{center}
Preprint, published as \emph{NeuroImage}, 89:345–357, 2014.\\
doi:10.1016/j.neuroimage.2013.11.043 
\end{center}

\paragraph{Abstract:}\label{abstract}

Multi-voxel pattern analysis (MVPA) is a fruitful and increasingly
popular complement to traditional univariate methods of analyzing
neuroimaging data. We propose to replace the standard `decoding'
approach to searchlight-based MVPA, measuring the performance of a
classifier by its accuracy, with a method based on the multivariate form
of the general linear model. Following the well-established methodology
of multivariate analysis of variance (MANOVA), we define a measure that
directly characterizes the structure of multi-voxel data, the pattern
distinctness $D$. Our measure is related to standard multivariate
statistics, but we apply cross-validation to obtain an unbiased estimate
of its population value, independent of the amount of data or its
partitioning into `training' and `test' sets. The estimate $\hat D$ can
therefore serve not only as a test statistic, but as an interpretable
measure of multivariate effect size. The pattern distinctness
generalizes the Mahalanobis distance to an arbitrary number of classes,
but also the case where there are no classes of trials because the
design is described by parametric regressors. It is defined for
arbitrary estimable contrasts, including main effects (pattern
differences) and interactions (pattern changes). In this way, our
approach makes the full analytical power of complex factorial designs
known from univariate fMRI analyses available to MVPA studies. Moreover,
we show how the results of a factorial analysis can be used to obtain a
measure of pattern stability, the equivalent of `cross-decoding'.

\paragraph{Keywords:}\label{keywords}

decoding; multivariate; multi-voxel pattern analysis; fMRI; MANOVA;
general linear model

\section{Introduction}\label{introduction}

In the last decade, traditional univariate methods for the analysis of
functional magnetic resonance imaging (fMRI) data based on the general
linear model (GLM) have increasingly been complemented by new methods
that try to access the information content of \emph{patterns} of
activation across a set of voxels. Variably termed multivariate (Haxby,
2012) or multi-voxel pattern analysis (Norman et al., 2006, MVPA),
information-based imaging (Kriegeskorte et al., 2006) or simply decoding
(Haynes and Rees, 2006), the new techniques have been applied to topics
as diverse as syntax and semantics (Mitchell et al., 2004), visual
perception (Kamitani and Tong, 2005), music and speech perception
(Abrams et al., 2011), and decision making (Kahnt et al., 2011). For a
recent review, see Tong and Pratte (2012).

Though other kinds of multivariate methods had been applied to
functional neuroimaging data before (Edelman et al., 1998; Friston et
al., 1993; Goutte et al., 1999; McIntosh et al., 1996; McKeown et al.,
1998; Worsley et al., 1997), the pioneering study of Haxby et al. (2001)
introduced something new by directly linking a multi-voxel activation
pattern to an experimental condition. Using a correlation-based
classifier, they achieved above-chance performance in predicting the
category of a visual stimulus from distributed cortical responses.
Extending this work, Cox and Savoy (2003) utilized linear discriminant
analysis (LDA) and support vector machines (SVMs) and showed that
classification performance tends to improve with the number of voxels
included. It was suggested that multi-voxel classification can be
sensitive to information predominantly represented at sub-voxel scales
(Kamitani and Tong, 2005), and may be used to access information not
consciously represented (Haynes and Rees, 2005a) as well as aspects of
subjective experience not determined by the stimulus (Haynes and Rees,
2005b). As an alternative to whole-brain or region-of-interest based
analyses, Kriegeskorte et al. (2006) introduced the `searchlight'
approach that performs multivariate analysis on sphere-shaped groups of
voxels centered on each brain voxel in turn. This method results in a
statistical map of local multivariate effects which can be seen as the
mass-multivariate counterpart to standard mass-univariate fMRI analysis.

The analysis of multi-voxel activation patterns in the `decoding'
literature relies almost exclusively on the use of \emph{classifiers}
(but see Haynes and Rees, 2005a; Kriegeskorte et al., 2006). Such a
classifier (cf. Pereira et al., 2009), often an SVM (Cortes and Vapnik,
1995), is `trained' to distinguish data vectors corresponding to two
classes of trials on part of the data, and its performance is assessed
or `tested' on the part not used for training. This is repeated such
that each part of the data set is once used for testing
(cross-validation), and the classification performance is quantified in
the form of an accuracy, the fraction of correctly classified test data
points. Nonlinear (kernel-based, see Boser et al., 1992) SVMs can be
used, but linear classifiers seem to be sufficient in most cases (Cox
and Savoy, 2003; Norman et al., 2006). This procedure can be applied to
raw fMRI volumes corresponding to single experimental trials, or to
estimates of parameters of run-wise GLMs. The latter approach has the
advantage that parameter estimates have clear attached class
memberships, while a single volume may represent a mixture of different
experimental conditions.

However, while classification algorithms have a large number of
important applications, their use for the purpose of searchlight-based
multi-voxel pattern analysis in basic neuroscience represents an
unnecessary detour. Classification is the right tool if the interest
actually lies in determining the group membership of single items. In
neuroscience the most important example of this are brain--computer
interfaces (Blankertz et al., 2011; Wolpaw and Wolpaw, 2012), where the
goal is e.g.\ to interpret the intention of a patient for every single
neural response, in order to initiate the correct action of a robot arm
or communication device. In most decoding studies in cognitive
neuroscience, however, the classification result is only considered in
the form of the \emph{accuracy} or another aggregate statistic, that is
then entered into statistical procedures in order to make an inference
about the neural processes underlying a cognitive operation. The
interest does not lie in the classification result per se, but in
whether classification can be performed better than chance.

Above-chance classification is possible if and only if the multivariate
distribution of the data is different between conditions. In this paper,
we therefore propose to avoid the detour through classification by
replacing the accuracy as the statistic used in searchlight-based MVPA
studies by a measure that directly quantifies the degree to which
multivariate distributions differ from each other: the \emph{pattern
distinctness} $D$. This measure generalizes the Mahalanobis distance
(used by Kriegeskorte et al., 2006) to the case of an arbitrary number
of classes of trials, and it can also be used with parametric
regressors. It is based on the multivariate version of the general
linear model (the MGLM) and therefore related to standard statistics
used in the multivariate analysis of variance (MANOVA; see Timm, 2002),
but cross-validation is applied in order to make it an unbiased
estimator of its population value.

The approach of cross-validated MANOVA to searchlight-based MVPA
proposed in this paper has a number of advantages over the
classification approach:\\-- The result of classification analysis
depends on the kind of algorithm chosen, its parameters, and on the way
the classifier is applied to the data (single trial vs run-wise
parameter estimates). By contrast, our method provides a unified
parameter-free framework based on a probabilistic model of the data,
which is the direct generalization of the standard univariate model for
fMRI data, the GLM.\\-- Classification accuracy as the numerical value
resulting from classifier-based MVPA does not have an interpretation
with respect to the underlying neuroimaging data, because its value does
not only depend on the `test' data being classified, but also on the
amount of training data used to construct it. By contrast the pattern
distinctness directly characterizes the data by measuring the amount of
multivariate variance accounted for by a specific effect, in relation to
the amount of error (or `noise') variance.\\-- Classifiers can only
assess how discriminable data corresponding to different experimental
conditions is. In an experimental design involving two or more factors,
it is often interesting to see whether the difference between the levels
of one factor is itself different across the levels of another factor:
an interaction analysis. While such an analysis can not be implemented
using classifiers, the framework of the MGLM is perfectly suited to
describe factorial designs and analyze arbitrary contrasts, including
main effects and interactions.

The measure of pattern distinctness is derived in the next section. The
third section shows the results of its application to a study on object
perception. Before concluding, we discuss the assumptions and possible
limitations of our proposed method, including its possible extension to
the case of whole-brain and region-of-interest based MVPA.

\section{Method}\label{method}

In the following we derive our measure of multivariate effect in a
sequence of steps. Starting with an information-theoretic motivation of
Mahalanobis distance, we generalize from the underlying normal
distribution model to the MGLM, and arrive at the algorithm of
cross-validated MANOVA (cvMANOVA). We investigate the properties of the
pattern distinctness $D$ theoretically and in a numerical simulation and
show that our measure can not only quantify pattern differences and
pattern changes, but also pattern stability, the equivalent of
`cross-decoding'.

\subsection{Mahalanobis distance}\label{mahalanobis-distance}

How well data points from two classes can on average be discriminated
depends on how different the underlying data distributions are. An
alternative to measuring the performance of a classifier is therefore to
model those distributions and to quantify their difference. The
assumption underlying standard univariate fMRI analyses (Friston et al.,
1995; Kiebel and Holmes, 2007) is that responses in different `classes'
are normally distributed, with different class means but the same
variance (homoscedasticity). The straightforward generalization of this
to the multivariate case is \[ \label{twoclass}
\vec y_i \dist \Dist N (\vec \mu_i, \Sigma),
\] where $\vec y_i$ is a $p$-dimensional activation vector (measured
BOLD signal values in $p$ different voxels) for classes $i = 1, 2$, and
$\vec \mu_i$ is the corresponding class mean activation pattern.
$\Sigma$ is a $p \times p$ covariance matrix describing the within-class
variance, and $\Dist N$ denotes the multivariate normal distribution.

\begin{figure}[p]
\centering
\includegraphics{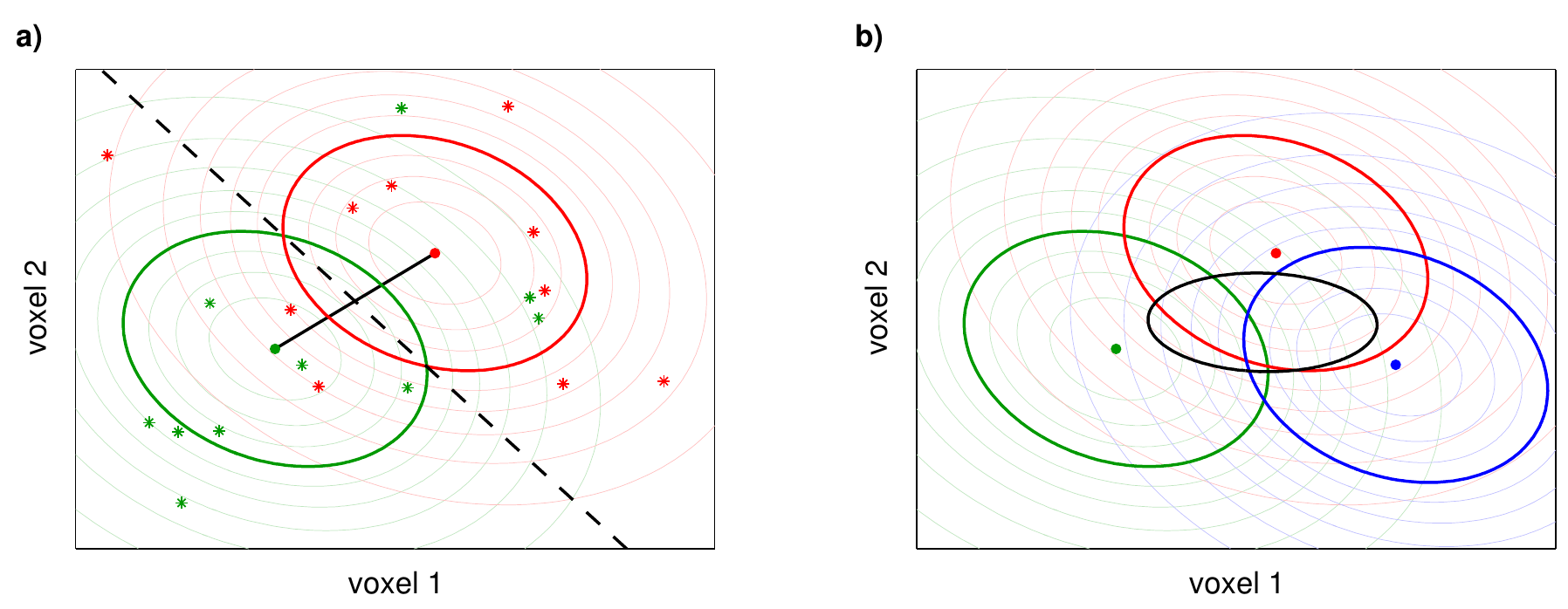}
\caption{\textbf{Classification and covariance structure.}
\ \textbf{(a)}\ Data points in voxel activation space that belong to two
different classes (red and green stars) can be discriminated using a
classification boundary (dashed line) if they come from different
distributions (densities indicated by contour lines); here an accuracy
of $70\,\%$ is achieved. Multivariate normal distributions as shown here
are defined by their expectation values (red and green center bullets)
and covariance structure (visualized by the strong red and green
ellipse-shaped $1\,\sigma$-lines). In this case, the discriminability of
the two distributions is completely characterized by the Mahalanobis
distance $\Delta$, the distance of the distribution centers (black line)
measured in standard deviations; here $\Delta = 1.5$. \ \textbf{(b)}\ The
discriminability of three or more classes is characterized by the
magnitude of the between-class covariance (black ellipse) compared to
the within-class covariance (colored ellipses). This magnitude can be
quantified e.g.\ by the diagonal diameter of the between-class covariance
ellipse measured in standard deviations of the within-class
distribution. (Note that the between-class covariance ellipse does not
correspond to the $1\,\sigma$-line of a multivariate normal
distribution, but describes the covariance structure of the three class
means treated as a sample.)}
\end{figure}

In general, the distinctness of two distributions can be quantified
using an information-theoretic measure, the Kullback--Leibler divergence
$D_\mathup{KL}$ (Kullback and Leibler, 1951). For two equal-variance
normal distributions (see Fig.\ 1a), this distinctness can be simply
expressed by the squared distance of the class means $\vec \mu_1$ and
$\vec \mu_2$, measured with respect to the within-class covariance
$\Sigma$, \[
2 \, D_\mathup{KL} = \  \ 
\Delta^2 = (\vec \mu_2 - \vec \mu_1)' \  \Sigma^{-1} \  (\vec \mu_2 - \vec \mu_1),
\] where $'$ denotes transposition. $\Delta$ is called \emph{Mahalanobis
distance} (Mahalanobis, 1936).\footnote{The same two-class model
  underlies linear discriminant analysis. Mahalanobis distance and LDA
  make complementary uses of the same distributional assumptions:
  quantifying the distinctness of the classes or assigning new data to
  them.}

The highest possible mean classification accuracy, achieved by a linear
classifier based on perfect knowledge of $\vec \mu_1$, $\vec \mu_2$, and
$\Sigma$ (see Hastie et al., 2009), can be derived theoretically; it is
\[
\alpha_\mathup{opt} = \Phi \left (\frac\Delta2 \right ),
\] where $\Phi$ denotes the cumulative distribution function of the
standard normal distribution. If the classifier has been trained on a
limited amount of data, imprecise estimation of the distribution
parameters leads to worse mean classification performance, but still the
mean accuracy $\alpha$ is a monotonically increasing function of
$\Delta$, starting from a chance level of $50\,\%$ and saturating for
large $\Delta$ towards $100\,\%$ (for approximate formulas see Wyman et
al., 1990).

In contrast to the expected accuracy $\alpha$ estimated by the
classifier-based approach, the equivalent quantity $\Delta$ directly
characterizes the multivariate data structure. Other than the accuracy,
its estimate does not depend on the internals of a particular
classification algorithm and it does not saturate for stronger effects.
Moreover, it is the multivariate generalization of a standard univariate
measure of effect size, Cohen's\ $d$ (Cohen, 1988).

\subsection{The Multivariate General Linear
Model}\label{the-multivariate-general-linear-model}

The multivariate normal distribution model for two-class data
\eqref{twoclass} is a special case of the multivariate general linear
model (MGLM), \[ \label{mglm}
Y = X B + \Xi.
\] The equation has the same form as for the GLM underlying univariate
fMRI analyses (Friston et al., 1995; Kiebel and Holmes, 2007), except
that here the measured data $Y$ is not a time series column vector but
of an $n \times p$ matrix specifying the signal at $n$ different time
points in $p$ voxels simultaneously. The design matrix $X$ specifies
time series (columns) for each of the $q$ regressors and remains
unchanged from the univariate case. Accordingly, the parameter matrix
$B$ is of size $q \times p$ and describes the strength of the
contribution of each of the $q$ regressors to the signal within each of
the $p$ voxels. Finally, each of the $n$ rows of the error term $\Xi$ is
a sample from a $p$-dimensional normal distribution,
$\Dist N (0, \Sigma)$. The following calculations are based on the
premise that the rows of $\Xi$ are mutually uncorrelated. Since fMRI
data are characterized by serial correlations, we assume that they have
been removed by a standard whitening procedure during preprocessing
(Glaser and Friston, 2007).

The two-class model \eqref{twoclass} can be written in the form of the
MGLM by including two regressors, such that $X$ contains a $1$ in the
first column for each data point belonging to class 1, a $1$ in the
second column for each data point belonging to class 2, and zeros
otherwise. The two rows of $B$ then correspond to vectors $\vec \mu_1$
and $\vec \mu_2$.

In GLM-based fMRI analyses, \emph{contrast matrices} $C$ are used to
select specific components out of the total space of effects that the
respective model can describe. Each contrast matrix defines a univariate
null hypothesis of the form $C' B = 0$, i.e.\ the statement that a GLM
parameter or a linear combination of parameters is zero, and forms the
basis for the computation of a $t$ or $F$ statistic used to test that
hypothesis. The same logic applies to the MGLM, only that now the
parameters in $B$ belonging to one regressor are vectors across $p$
voxels, and the specified null hypotheses are multivariate null
hypotheses. If contrasts are used to select partial effects
corresponding to single factors and interactions in a factorial design,
this leads to a decomposition or `analysis' of variance (ANOVA) in the
case of the GLM and to a multivariate analysis of variance, MANOVA, in
the case of the MGLM.

The formal development is again identical in both cases (cf. Kiebel and
Holmes, 2007).\footnote{Notation: $A^-$, Moore--Penrose pseudo-inverse
  of a matrix $A$. $\hat B$, estimate of a model parameter $B$.
  $\langle X \rangle$, expectation value of a random variable $X$.} The
contrast matrix $C$ is used to separate the parameters of the full model
into two parts, \[
B = B_0 + B_\Delta
\] with \[  \label{extract}
B_\Delta = C'^- C' B = C C^- B
\quad \text{and} \quad
B_0 = B - B_\Delta.
\] $B_0$ describes a reduced model corresponding to the null hypothesis
$C' B = 0$, and $B_\Delta$ a possible deviation from this reduced model.
In the case of two classes of trials modeled by two regressors and a
contrast $C = (-1 \  1)'$, $B_\Delta = \vec \mu_2 - \vec \mu_1$ describes
the difference of the multivariate activation patterns between these
classes.

The \emph{between-class covariance}, i.e.\ the part of the total
covariance of the data that is accounted for by the class difference, is
given by \[
\frac1n \, B_\Delta' X' X B_\Delta,
\] while the within-class covariance is identical to the \emph{error
covariance}, \[
\frac1n \, \left \langle \Xi' \Xi \right \rangle = \Sigma.
\] The size of the multivariate effect is the magnitude of the
between-class covariance compared to the within-class covariance
(Fig.\ 1b). Since both are $p \times p$ matrices, there are different
ways to express this comparison in a single number; we choose
\[ \label{trace}
D = \tr \left ( \frac1n B_\Delta' X' X B_\Delta \ \  \Sigma^{-1} \right ),
\] because in the two-class case it has a simple relationship to the
Mahalanobis distance: \[
D = \frac14 \Delta^2 \ \cdot\  \frac{n_1 + n_2}{n} \ ,
\] where $n_1$ and $n_2$ are the number of data points for class 1 and
2, respectively. We call this measure \emph{pattern distinctness}
because it quantifies how distinct different multivariate patterns
appear relative to the uncertainty induced by the error.

But the MGLM and the pattern distinctness $D$ are more than just a
different way to consider the two-class model. The MGLM can be used with
regressors of arbitrary form, i.e.\ they can be simple indicators for two
or more classes of trials, indicator variables convolved with the
hemodynamic response function, or general parametric regressors where
there are no classes.

This allows to apply the MGLM directly to the measured fMRI signal,
while a classifier might have to be applied to (run-wise) GLM parameter
estimates $\hat B$ in order to provide clear `class identities' for the
data points being entered. Moreover, modeling directly the data allows
to estimate the `within-class' covariance, which is actually the error
covariance $\Sigma$, on a \emph{volume-by-volume basis}, leading to a
precise estimate.

Additionally, $D$ is defined for a contrast $C$ specifying a simple
two-class comparison, but also for multi-class comparisons, for main
effects in a factorial design, or for arbitrary (estimable) model
parameter combinations. While some of these comparisons may be emulated
to some degree using classifiers, e.g.\ multi-class comparison by
multiple pairwise classification, this is not possible in the case of an
interaction of several experimental factors. In short, by using the MGLM
the full analytical power of \emph{complex factorial designs} becomes
available to multi-voxel pattern analysis.

Just as the Mahalanobis distance $\Delta$ characterizes the multivariate
data structure in the two-class case, the pattern distinctness $D$
fulfills this function for the general case of a partial effect defined
by a contrast $C$ within an experimental design described by a design
matrix $X$. In contrast to the classification accuracy, $D(C)$ has a
clear interpretation as it quantifies the amount of multivariate
variance (Cohen, 1982) explained by the effect, measured in units of the
error variance.

\subsection{Cross-validated MANOVA}\label{cross-validated-manova}

The Maximum-Likelihood fit of an MGLM to a given data set is achieved by
Least Squares: \[ \label{step1}
\hat B = X^- \  Y
\quad \text{and} \quad
\hat \Xi = Y - X \hat B.
\] The straightforward estimate of the pattern distinctness $D$ based on
this fit is identical to the \emph{Bartlett--Lawley--Hotelling trace}:
\[
T_\mathup{BLH} = \tr ( H \  E^{-1} ).
\] Here \[
H = \hat B_\Delta ' X' X \hat B_\Delta
\quad \text{and} \quad
E = \hat \Xi' \hat \Xi
\] are the hypothesis and error matrices of sums of squares and
cross-products. In the univariate case ($p = 1$), $H$ and $E$ are simple
sums of squares and the ratio $F = H / E \cdot f_E / f_H$ gives the
ANOVA statistic with degrees of freedom $f_H = \rk X C$ and
$f_E = n - \rk X$.

Along with Wilks' $\Lambda$ (cf. Haynes and Rees, 2005a), the
Bartlett--Lawley--Hotelling trace $T_\mathup{BLH}$ is one of the
standard test statistics used in multivariate statistics (MANOVA,
canonical correlation analysis, etc.; see Timm, 2002). As an estimator
of $D$, however, it is severely biased. An intuitive explanation is that
$T_\mathup{BLH}$ is a (generalized, squared) distance. Since the
estimates cannot be negative, if $D$ is zero or small, estimation errors
mostly increase the estimate. Or from a model evaluation perspective,
$H$ can be seen as comparing (`correlating') the estimate of the
contrast effect on the data, $X \hat B_\Delta$, with itself, leading to
an overestimation of the effect.

As in the case of estimating accuracies, this problem can be remedied
using \emph{cross-validation} (Hastie et al., 2009). For a data set
consisting of $m$ runs, $k = 1 \ldots m$, we perform a
`leave-one-run-out' cross-validation. The estimate of $D$ in the $l$th
cross-validation fold, \[ \label{step3}
\hat D_l = \tr ( H_l \  E_l ^{-1} ),
\] combines $\hat B_\Delta$ computed from data of the $l$th (`left out'
or `test') run on the right-hand side with its values from the other
(`training') runs on the left-hand side: \[ \label{step2}
H_l = \sum_{k \neq l} \left \{ \hat B_\Delta' \right \}_k \ 
\left \{ X' X \hat B_\Delta \right \}_l
\quad \text{and} \quad
E_l = \sum_{k \neq l} \left \{ \hat \Xi' \hat \Xi \right \}_k,
\] where braces with subscript indicate that the expression has to be
evaluated using data from the respective run. The complete
cross-validated estimate of $D$ is then the mean of the per-fold
estimates $\hat D_l$. After additionally correcting a multiplicative
bias of $E^{-1}$ in estimating $(n \Sigma)^{-1}$, the final unbiased
estimator of the pattern distinctness $D$ becomes (see Appendix A)
\[ \label{step4}
\hat D
= \frac{(m - 1) \, f_E - p - 1}{(m - 1) \, n}
\cdot \frac1m \sum_{l = 1}^m \hat D_l.
\]

While the Bartlett--Lawley--Hotelling trace $T_\mathup{BLH}$ is useful
as a test statistic to determine whether the observed effect is
significantly different from zero, its cross-validated version $\hat D$
can serve the same purpose, but additionally provides an unbiased
estimate of the \emph{size of the effect}. Such a characterization of
the effect independent of the amount of data available or its
partitioning into `training' and `test' sets enables interpretation and
potentially facilitates cross-study comparisons and power analyses (cf.
Cohen, 1988). However, the estimated effect size $D$ itself does of
course depend on the specific design matrix $X$ used (e.g.\ number of
trials, type of regressors), and may depend on the number of voxels $p$
included in the analysis.

A small modification is advisable for the purpose of topologically
specific inference using a statistical parametric map of
$\hat D$-values. Since searchlight spheres centered near the boundaries
of the brain mask contain fewer voxels and the variance of the null
distribution of $\hat D$ is approximately proportional to $p$ (see
Appendix B), the map has an inhomogeneous null variance. We therefore
recommend to use statistical parametric maps of the \emph{standardized}
pattern distinctness \[ \label{step5}
\hat D_\mathrm s = \frac1{\sqrt{p}} \, \hat D,
\] and to additionally report peak or mean values of $\hat D$.

\subsection{Simulation}\label{simulation}

The statistical properties of $\hat D$ were investigated using
artificial data sets approximately simulating an fMRI experiment. It
comprised $m = 4$ runs of $n = 512$ volumes each. There were
$n_1 = n_2 = 16$ trials per run for each condition, each lasting for one
volume, such that the design matrix consisted of indicator variables for
the two conditions and a constant regressor. Data were generated
according to the MGLM equation \eqref{mglm} with serially uncorrelated
errors, i.e.\ the simulation refers to data after the whitening
preprocessing step. The error covariance was set to $\Sigma = \I$
without loss of generality (cf.\ Appendix A). The contrast considered was
$C = (-1 \  1 \  0)'$. For each choice of simulation parameters (true
effect size $D$ and dimensionality $p$), $10,000$ data sets were
generated.

For each artificial data set, additionally the empirical accuracy $a$
was determined for classification of run-wise parameter estimates
$\hat \beta_{1\cdot}$ and $\hat \beta_{2\cdot}$ ($p$-dimensional
vectors) as well as for classification of single-trial volumes, in each
case using a linear soft-margin support vector classifier (Cortes and
Vapnik, 1995), in the implementation of the LIBSVM library (Chang and
Lin, 2011).

\begin{figure}[p]
\centering
\includegraphics{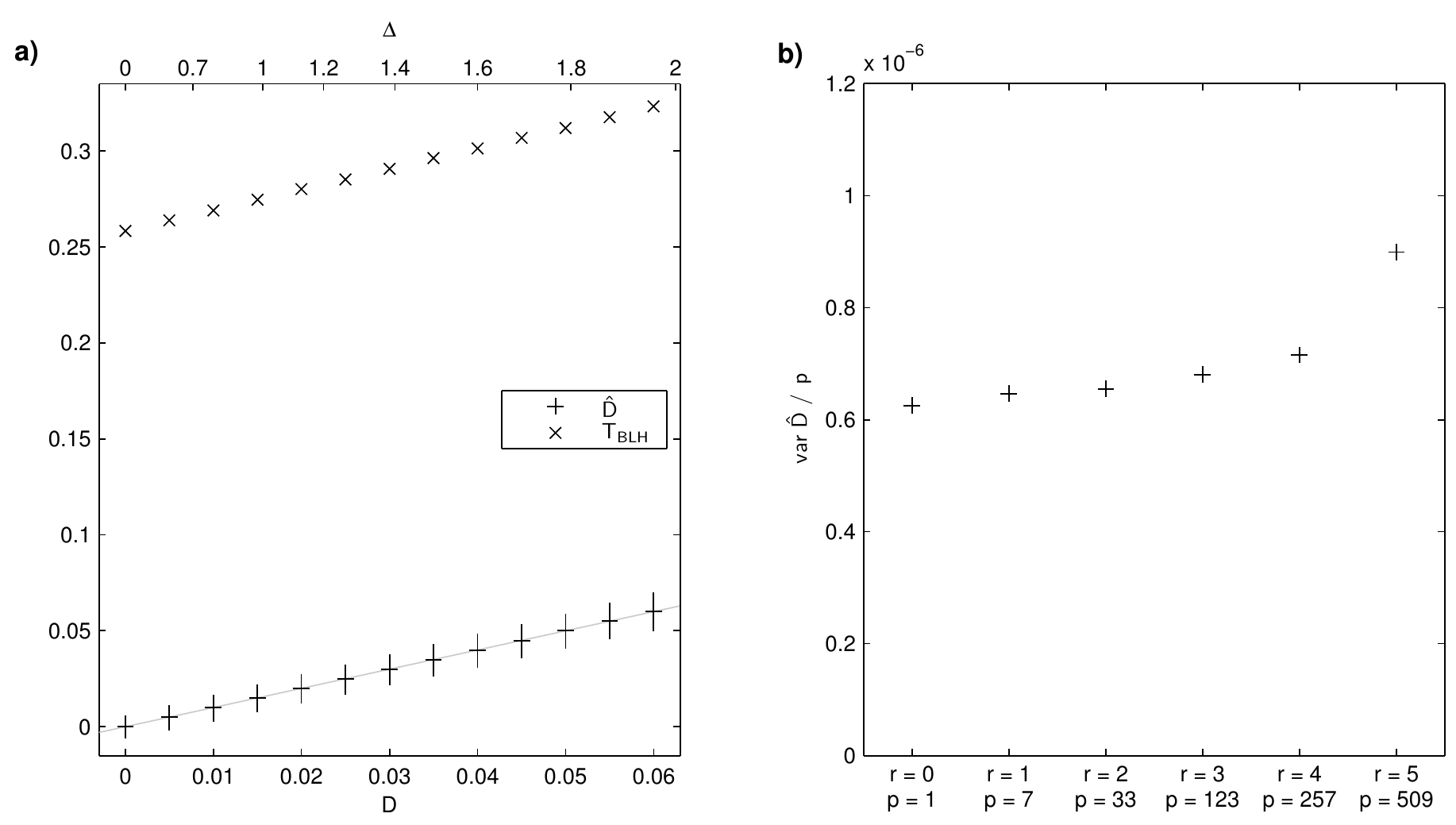}
\caption{\textbf{Simulation results: Statistical properties.}
\ \textbf{(a)}\ The cross-validated and naïve estimates of $D$, $\hat D$
and $T_\mathup{BLH}$, as a function of $D$ for $p = 123$ voxels. Values
of $\hat D$ are shown with a marker ($+$) indicating the mean and a
vertical line for the interquartile range of the sampling distribution,
values of $T_\mathup{BLH}$ only with a marker ($\times$). The identity
line $\hat D = D$ is shown in gray. For reference, the upper horizontal
scale gives the corresponding values of the Mahalanobis distance
$\Delta$. The plot confirms that $\hat D$ is an unbiased estimator of
$D$, while the Bartlett--Lawley--Hotelling trace $T_\mathup{BLH}$ is
strongly biased. \ \textbf{(b)}\ The ratio $\var \hat D / p$ for $D = 0$,
as a function of the number of voxels $p$, corresponding to searchlight
radii $r$ of 0 to 5. The plot supports the theoretical expectation that
the sampling variance of $\hat D$ under $\Hnull$ is approximately
proportional to $p$ for moderately large numbers of voxels.}
\end{figure}

Fig.\ 2a shows $\hat D$ and $T_\mathup{BLH}$ as a function of $D$ for
$p = 123$ voxels, corresponding to a searchlight radius of 3 voxels. The
simulation demonstrates the extreme estimation bias of $T_\mathup{BLH}$,
and confirms the theoretical result that $\hat D$ is an unbiased
estimator of $D$.

Fig.\ 2b shows the sampling variance of $\hat D$ under the condition that
there is no true effect ($D = 0$), as a function of $p$. It demonstrates
that the ratio $\var \hat D / p$ is approximately constant for small to
moderate values of $p$. This supports the theoretical expectation of
proportionality between null variance and dimensionality, which means
that the standardized pattern distinctness
$\hat D_\mathrm s = \tfrac1{\sqrt{p}} \hat D$ is a suitable statistic to
construct statistical parametric maps.

\begin{figure}[p]
\centering
\includegraphics{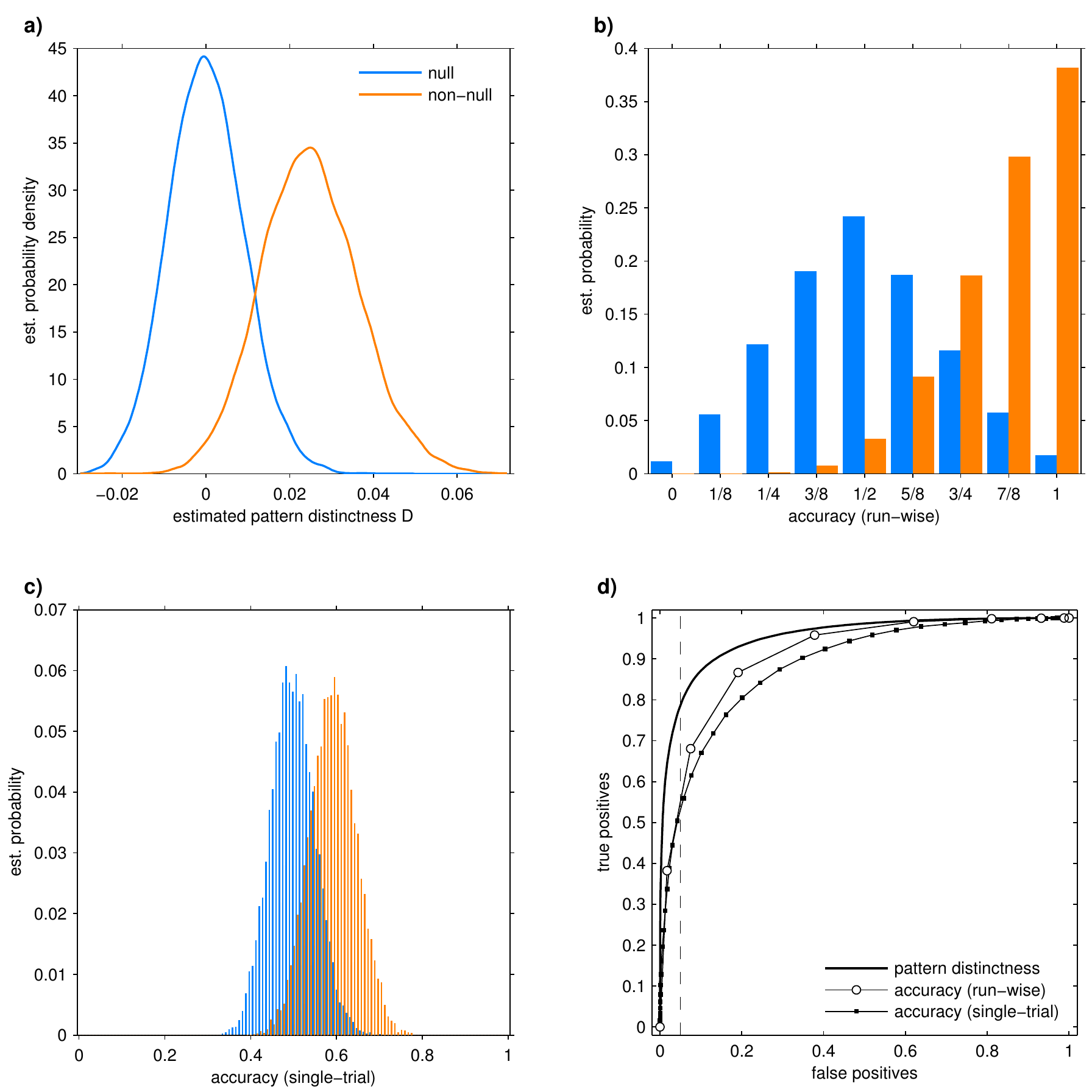}
\caption{\textbf{Simulation results: Effect detection.}
\ \textbf{(a)}\ Sampling distribution of the estimated pattern
distinctness $\hat D$ for true values $D = 0$ (null effect, blue) and
$D = 0.025$ (non-null effect, orange), for $p = 123$ voxels. (Gaussian
kernel density estimate with bandwidth $0.0014$) \ \textbf{(b)}\ Sampling
distribution of the empirical accuracy for classification of run-wise
parameter estimates, for the same true effect sizes. Based on a
classification of two data points in each of four runs, $a$ can only
take on nine different values. \ \textbf{(c)}\ Sampling distribution of
the empirical accuracy for classification of single-trial volumes, for
the same true effect sizes. Here, $a$ can take on 129 different values,
of which 60 actually occur. \ \textbf{(d)}\ Receiver operating
characteristic for distinguishing the non-null from the null effect
based on the observed value of $\hat D$ (strong line), run-wise
classification accuracy (circle-line), and single-trial classification
accuracy (dot-line, respectively. The diagram illustrates the trade-off
between the fraction of samples falsely, to the fraction of samples
correctly determined to come from the non-null (orange) distribution,
for varying thresholds. At a fixed false positive rate (`type I error')
of $0.05$ (vertical dashed line), the pattern distinctness achieves a
true positive rate (`power') of $0.79$ compared to an interpolated true
positive rate for run-wise accuracy of $0.55$ and for single-trial
accuracy of $0.53$.}
\end{figure}

The usefulness of the pattern distinctness $\hat D$ and the empirical
accuracy $a$ as a statistic to distinguish data sets with a true
underlying effect from those with a null effect is investigated in
Fig.\ 3. The chosen non-null effect of $D = 0.025$ can be considered
`large' if the corresponding $\Delta = 1.26$ is compared to Cohen's
recommendations for $d$. Accordingly, the sampling distributions of
$\hat D$ for null and non-null effect (Fig.\ 3a) overlap only moderately.
The run-wise classification accuracy (Fig.\ 3b) is not able to resolve
these two effects as well because the range of possible values is
limited ($[0, 1]$) and completely covered by the null distribution. The
increased range of possible values for single-trial classification
(Fig.\ 3c) mildens the problem, but still the sampling distributions for
null- and non-null effect appear to overlap more than for $\hat D$.

The same observation is presented in a more precise way in Fig.\ 3d in
the form of a `receiver operating characteristic' diagram (ROC, see also
Kriegeskorte et al., 2006). If a data set is determined to show an
effect based on comparison with a threshold, the trade-off between false
and true positive rate can be adjusted by choosing the threshold value.
The ROC curves show that in this simulation $\hat D$ achieves a higher
true positive rate than both run-wise and single-trial classification
accuracies for arbitrary given false positive rate. Though this result
suggests that hypothesis tests based on $\hat D$ may be more powerful
than tests based on $a$, their relative performance will depend on the
exact structure of the data set and the chosen comparison; see Sec.\ 3
for an example where $a$ shows statistically stronger effects than
$\hat D$.

\subsection{Pattern stability}\label{pattern-stability}

A common variation of classifier-based pattern analysis is
`cross-decoding' (cf. Haynes and Rees, 2005b), i.e.\ training a
classifier on one pair of classes and testing it on another pair of
classes. This method is typically applied in a factorial design, where
the two pairs of classes refer to the same factor but under different
levels of a second factor (see next section for an example).
Cross-decoding is therefore the complement of an interaction analysis:
while the latter attempts to show that a pattern difference changes
under an additional manipulation, cross-decoding attempts to show that
it remains \emph{stable}. This approach can be translated into the
framework presented here; since our measure of multivariate effect is
based on `correlating' parameter estimate differences $\hat B_\Delta$
between `training' and `test' runs \eqref{step2}, it is also possible to
apply it to estimate differences computed for different contrasts.

However, the explicit computation of such a modified form of the pattern
distinctness $D$ is not necessary. The degree of stability of a
multivariate effect $\E$ across the levels of another experimental
factor $\A$ can be quantified by combining the multivariate measures of
the effect $\E$ and the interaction $\E \times \A$. Under the hypothesis
of maximal inconsistency (mutual orthogonality) of patterns across the
$L$ levels of $\A$ it holds \[
D(\E) = \frac1{L - 1} \  D(\E \times \A)
\] (see Appendix C). The difference of the two sides of the equation
therefore quantifies the degree of pattern stability. Adopting the
symbol for the set complement (`$\setminus$') to denote the complement
of an interaction (`$\times$'), we can define \[ \label{deltad}
D(\E \setminus \A) = D(\E) - \frac1{L - 1} \  D(\E \times \A)
\] as the multivariate \emph{measure of pattern stability}. It is $0$
for maximal pattern inconsistency and attains its maximum value of
$D(\E)$ for zero interaction, i.e.\ maximum pattern stability.

The expression can be interpreted such that the total amount of variance
accounted for by the effect $\E$ is reduced by that part that is
inconsistent across the levels of $\A$, quantified by the strength of
the interaction. Applied to estimates $\hat D$ of the multivariate
effects, it can be used as a test statistic to reject the null
hypothesis of maximal pattern inconsistency and thereby provide evidence
for pattern stability.

\section{Application}\label{application}

\begin{figure}[p]
\centering
\includegraphics{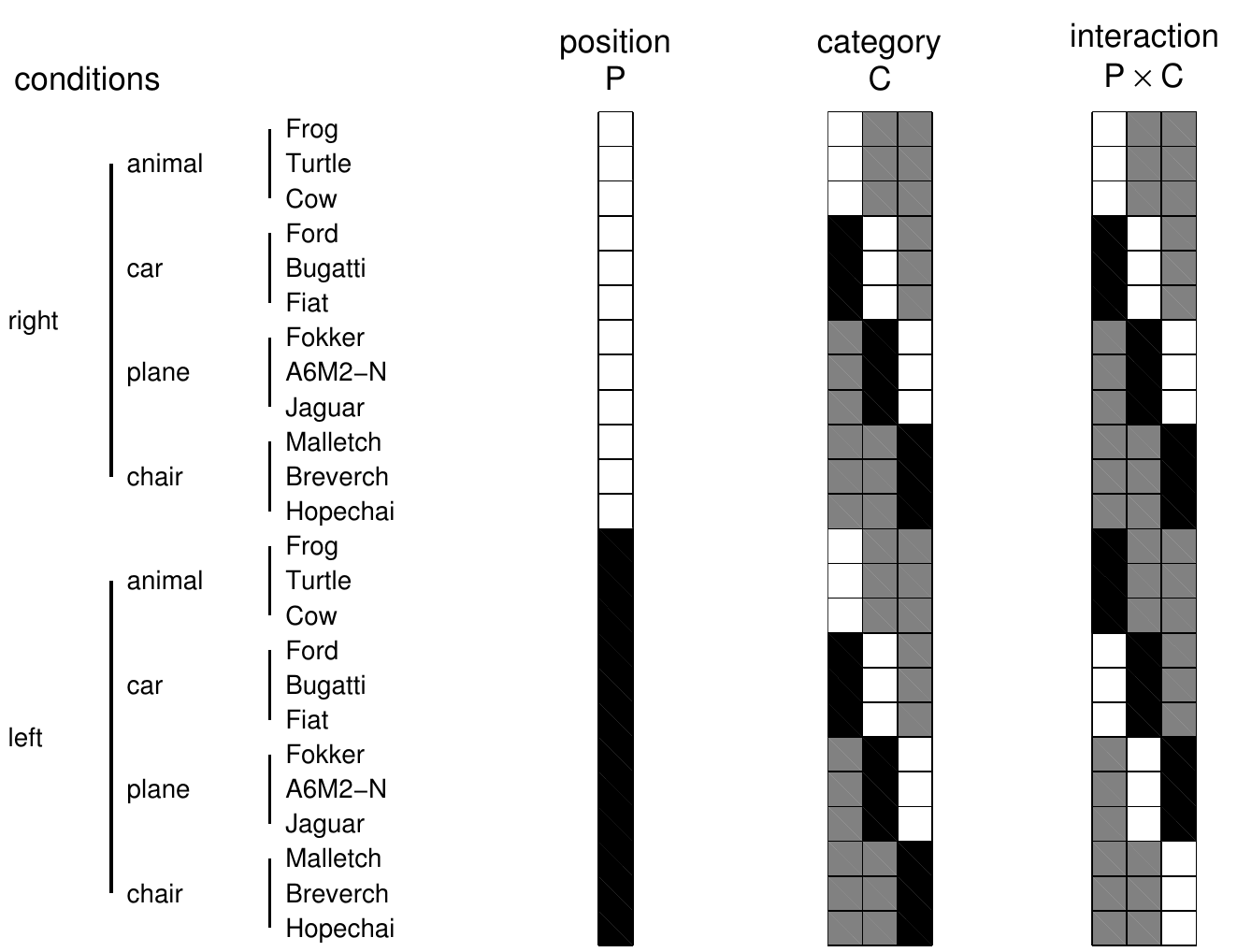}
\caption{\textbf{Experimental conditions and contrasts.} \ Renderings of
twelve different objects belonging to four different categories were
shown left and right of a fixation dot. The 24 resulting conditions were
examined in the form of three contrasts corresponding to the main
effects of position (P) and category (C), and their interaction
(P\ $\times$\ C). The associated contrast matrices $C$ are shown in
grayscale, where white stands for a value of $1$, black for $-1$, and
gray for $0$.}
\end{figure}

In order to demonstrate the use of our measure of pattern distinctness,
we reanalyzed the data of Cichy et al. (2011). In this study, renderings
of three-dimensional object meshes were presented to subjects either to
the left or the right of a central fixation dot. The objects belonged to
four different categories with three exemplars for each category,
resulting in 24 experimental conditions (see Fig.\ 4). Images were
presented in mini-blocks of six different views of the same object, at a
rate of one view per second. There were four mini-blocks per condition
in each of the five experimental runs. Data were acquired in 412 fMRI
volumes per run at a TR of $2\,\mathrm s$, with a field of view covering
the ventral visual cortex at an isotropic resolution of
$2\,\mathrm{mm}$. Volumes were slice-time corrected, realigned and
normalized to the MNI template. Thirteen subjects participated in the
experiment, but the data of one were discarded because of strong head
movements. The data of each subject were modeled with one design matrix
per run, each comprising regressors for the 24 conditions. Regressors
were composed of canonical HRFs time-locked to each image presentation.
Preprocessing, construction of the design matrix, and removal of serial
correlations were performed in SPM8
(\url{http://www.fil.ion.ucl.ac.uk/spm/}), further analysis in custom
SPM-based MATLAB code (The Mathworks, Natick).

The experiment has a three-way design with the factors position (P:
right, left), category (C: animal, car, plane, chair), and exemplar (E),
where the factor E is nested within the factor C. We investigated the
main effects of P and C, as well as the interaction P\ $\times$\ C; the
corresponding contrast matrices are shown in Fig.\ 4.

Using a searchlight of radius 4 voxels ($p = 257$ voxels), for each
searchlight position we computed MGLM parameter estimates $\hat B$ and
residuals $\hat \Xi$ \eqref{step1}. The respective contrast was applied
to obtain $\hat B_\Delta = C C^- \hat B$. Performing a leave-one-run-out
cross-validation, for each fold $l = 1 \ldots m$ hypothesis and error
matrices $H_l$ and $E_l$ were computed \eqref{step2} and from them the
fold-wise estimate of pattern distinctness $\hat D_l$ \eqref{step3}. The
per-fold estimates were then combined into the final unbiased estimate
of pattern distinctness $\hat D$ \eqref{step4}. The result was converted
into a statistical parametric map, SPM$\{\hat D_\mathrm s\}$, of
standardized pattern distinctness \eqref{step5}. Maps computed for each
subject and contrast separately were subsequently smoothed with a
Gaussian kernel of $6\,\mathrm{mm}$ FWHM.

The application of standard second-level statistics including $t$-tests
to cross-validated MVPA measures has been discussed critically by
Stelzer et al. (2013). We instead followed their approach to perform a
group-level permutation test by combining single-subject permutation
values selected independently in each subject. These single-subject
permutations were generated by a sign-permutation procedure adapted for
cross-validated MANOVA, described in App.\ D. With $m = 5$ runs, there
were $2^4 = 16$ single-subject permutations and
$16^{12} = 2.8\cdot10^{14}$ combined permutations, out of which
$100,000$ were randomly selected. The group-level test statistic was the
standardized pattern distinctness $\hat D_\mathrm s$ averaged across
subjects. Statistical results were corrected for multiple comparisons at
the voxel level by using the permutation distribution of the maximum
statistic across voxels (Nichols and Holmes, 2002).

\begin{figure}[p]
\centering
\includegraphics{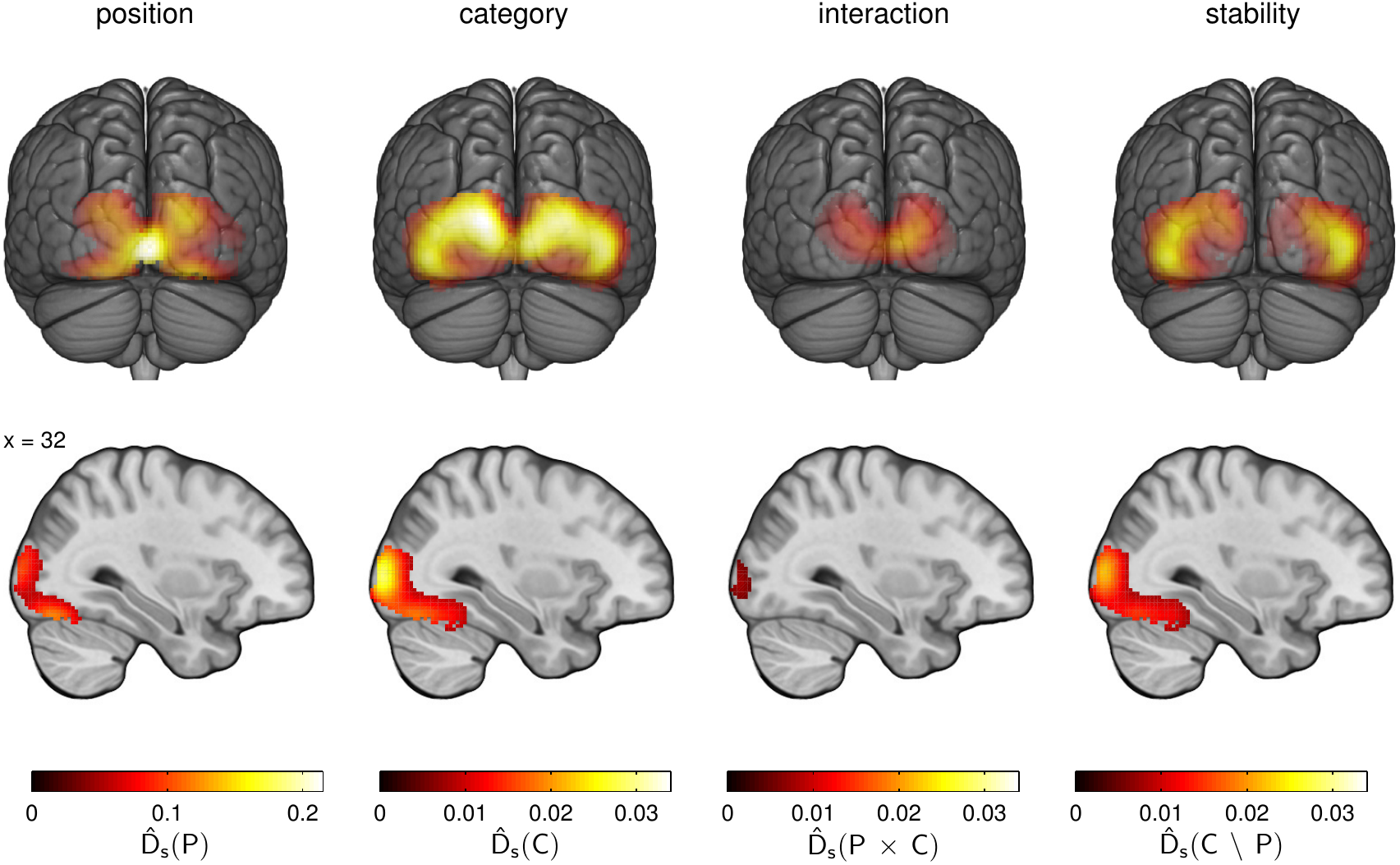}
\caption{\textbf{Analysis results.} \ The results of cvMANOVA in the form
of statistical parametric maps of the standardized pattern distinctness
$\hat D_\mathrm s$, averaged across twelve subjects, for four different
multivariate effects. The maps are shown on a 3D rendering of the
ICBM152 template brain and on a sagittal slice at $x = 32$. The
highlighted areas are those where the multivariate effect was
statistically significant at a level of $P \leq 0.05$, corrected for
multiple comparisons. The multivariate main effect of position (P) is
strongest over primary visual cortex, while the main effect of category
(C) extends from middle and superior occipital gyrus into the fusiform
gyrus. The multivariate interaction of these two factors (P\ $\times$\ C)
is again mainly confined to primary visual areas. In addition to the
three standard contrasts (see Fig.\ 4), the fourth column shows the
difference in effect size between the main effect of category and the
interaction,
$\hat D_\mathrm s (\mathrm C \setminus \mathrm P) = \hat D_\mathrm s (\mathrm C) - \hat D_\mathrm s (\mathrm P \times \mathrm C)$.
This difference quantifies the degree of stability of the patterns
representing category information under a change of presentation
position, and thereby indicates the presence of position-invariant
category information (equivalent to cross-decoding). Note that a
different color scale had to be used for the leftmost plot because of an
overall much stronger effect of position.}
\end{figure}

The results are shown in Fig.\ 5. The pattern distinctness for the three
computed contrasts obtains values significantly different from zero over
large areas of the visual cortex, with maxima exhausting the number of
computed permutations, $P \leq 10^{-5}$ (corr.). The same holds for the
measure of pattern stability
$\hat D_\mathrm s (\mathrm C \setminus \mathrm P) = \hat D_\mathrm s (\mathrm C) - \hat D_\mathrm s (\mathrm P \times \mathrm C)$
which quantifies the degree of stability of category-specific patterns
across the two positions \eqref{deltad}, i.e.\ indicates the presence of
position-invariant category information. The effect sizes reach maxima
of $\hat D = \sqrt{p} \, \hat D_\mathrm s = 3.45$ for the main effect of
position in primary visual cortex, for category of $\hat D = 0.55$
bilaterally in middle and superior occipital gyrus, and for the
interaction P\ $\times$\ C of $\hat D = 0.35$ again in V1. For the pattern
stability C\ $\setminus$\ P values of up to $\hat D = 0.45$ are reached
bilaterally over middle and inferior occipital gyrus as well as fusiform
gyrus, consistent with the location of the lateral occipital complex
identified by Cichy et al. (2011).

To compare cross-validated MANOVA with established analysis methods, we
also computed classification accuracies and mass-univariate statistical
parametric maps. However, please note that classification accuracy is
not defined for an interaction effect and there is no analogue of
`pattern stability' for univariate analysis. Cross-validated accuracies
were determined using an SVM to classify run-wise parameter estimates.
Accuracy maps were smoothed with a Gaussian kernel of $6\,\mathrm{mm}$
FWHM before entering statistical assessment, which was based on a
permutation procedure analogous to that applied to $\hat D$, exchanging
labels between conditions. Statistical significance of accuracies
averaged across subjects was again assessed by comparing actual values
with the permutation distribution of the maximum statistic, estimated by
$100,000$ times randomly combining single-subject permutations.
Additionally, we computed the univariate second-level SPM$\{F\}$ applied
to first-level contrast images (smoothed as above), statistically
assessed in SPM8 by standard familywise error correction based on random
field theory.

\begin{figure}[p]
\centering
\includegraphics{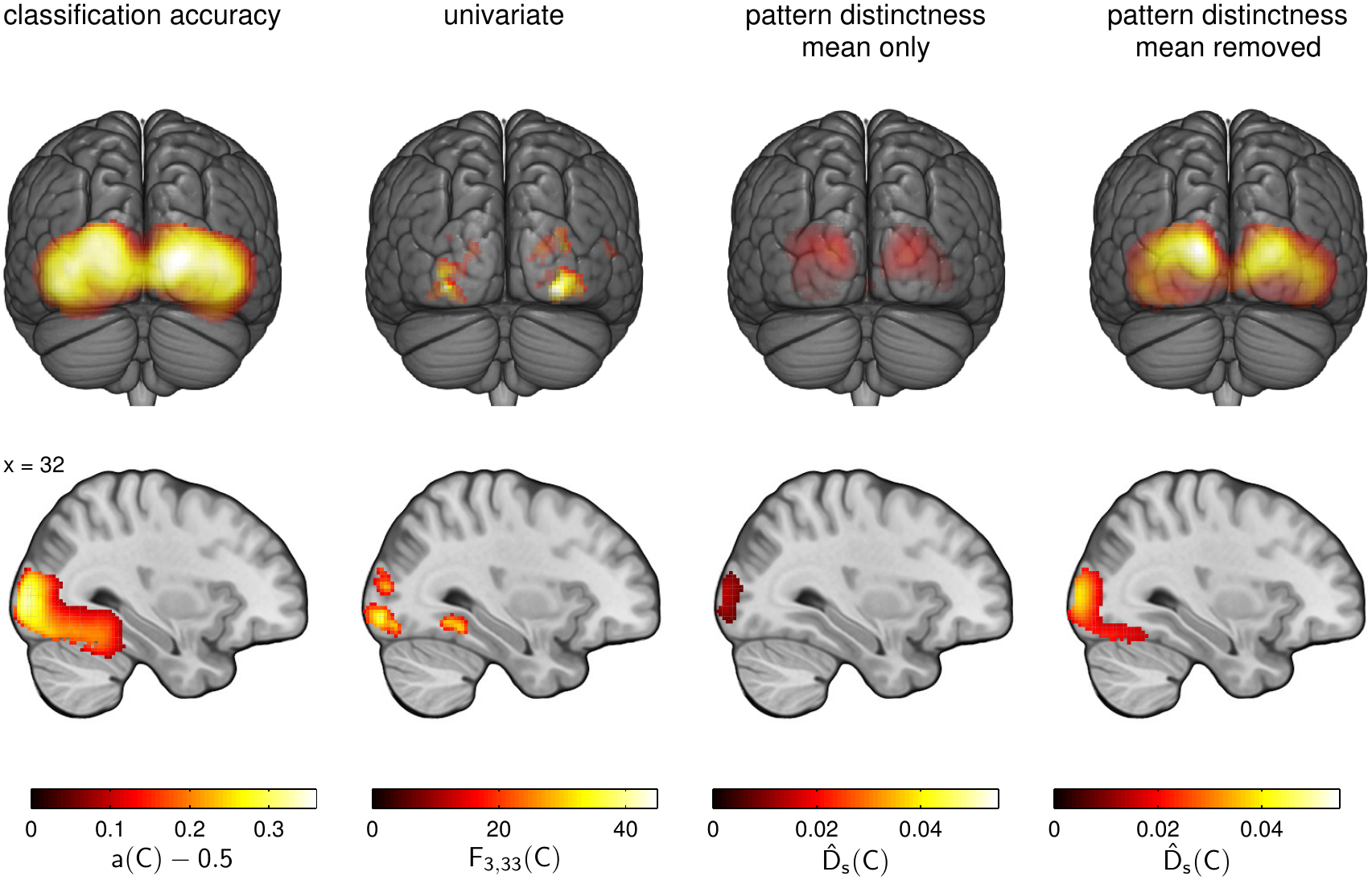}
\caption{\textbf{Comparison of analysis approaches.} \ Analysis of the
`category' effect with four different measures: cross-validated
above-chance classification accuracy averaged across subjects,
second-level univariate ANOVA F-value, average pattern distinctness
computed on searchlight-wise mean only, and average pattern distinctness
computed on mean-free data. The highlighted areas are those where the
observed effect was statistically significant at a level of
$P \leq 0.05$, corrected for multiple comparisons. Color scales are
adapted for each measure to cover the range from 0 to the maximum across
the brain (across both analyses for $\hat D_s$).}
\end{figure}

The results for `category' are shown in Fig.\ 6. The classification
analysis gives a similar picture of the localization of category
information as cvMANOVA. In contrast to the simulation results of Sec.
2.4, statistical power of the test based on the classification accuracy
$a$ appears to be higher than that based on the pattern distinctness
$\hat D$. Univariate effects are observed in areas consistent with the
multivariate analyses but turn out to be much weaker. To ensure that
this difference is not just due to different statistical methods, we
repeated the cross-validated MANOVA, but set all parameter estimates
within each searchlight to their mean before multivariate analysis. The
results (Fig.\ 6) are much weaker and especially fail to uncover category
effects in fusiform gyrus. By contrast, cvMANOVA computed on parameter
estimates from which the mean was removed for each searchlight location
separately give only slightly weaker results than the original analysis,
indicating that the observed effect has a predominantly non-univariate
character.

\section{Discussion}\label{discussion}

In the following we discuss the two critical assumptions underlying the
method of cross-validated MANOVA: normally distributed errors and
linearity of the multivariate model; as well as a possible limitation of
its applicability due to an insufficient amount of data.

\subsection{Nonnormality}\label{nonnormality}

As stated above, our data model assumes multivariate normally
distributed additive errors $\Xi$, which is the natural multivariate
generalization of the assumption underlying standard univariate fMRI
analyses. The theoretical justification for the normality assumption is
that the error term captures all those aspects of the operation of the
brain and the MR scanner which do not systematically occur in the
experiment and therefore cannot be modeled explicitly. Because these
processes are likely to be high-dimensional and their many small
contributions add up, according to the central limit theorem the error
can be expected to be normally distributed. Since this theorem
generalizes to the multivariate case (see Timm, 2002), the justification
also holds for the MGLM. In practice, residuals of properly constructed
models in fMRI do appear to be approximately normally distributed
(Kruggel and Cramon, 1999), and univariate analyses which depend on this
assumption have proven to be successful and reliable.

A within-class multivariate normal distribution is also the model
underlying the standard parametric approach to classification, LDA,
which was successfully used in several early MVPA studies (Carlson et
al., 2003; Cox and Savoy, 2003; Haynes and Rees, 2005a, 2005b). However,
as Hastie et al. (2009) state, optimal performance of linear classifiers
is not bound to multivariate normality; it extends to the family of
distributions that are characterized by a mean pattern $\vec \mu$, a
covariance matrix $\Sigma$ and a monotonically decreasing function
$g(d)$ such that the density is \[
p(\vec x) \propto g \left( (\vec x - \vec \mu)' \  \Sigma^{-1} \  (\vec x - \vec \mu) \right ),
\] the \emph{elliptical distributions} (see Fang et al., 1990). Since in
this family of distributions the probability of a data point is a
function of the Mahalanobis distance from the distribution center, the
argument translates to our measure: Quantifying the distinctness of
distributions using the Mahalanobis distance and our generalization $D$
is still appropriate for distributions heavier- or thinner-tailed than
the multivariate normal, as long as they fall within the elliptical
family.

A more serious concern is that deviations from model assumptions can
induce errors in statistical inference, especially an increased
probability of false positives (type I error rate), but also decreased
power. For this reason we recommend to base hypothesis tests using
$\hat D$ on nonparametric procedures, especially permutation tests
(Good, 2005; Nichols and Holmes, 2002), or at least to use nonparametric
methods to back up the results of parametric methods.

\subsection{Nonlinearity}\label{nonlinearity}

An advantage of classifiers might be seen in the fact that kernel-based
methods (Boser et al., 1992) allow for nonlinear classification.
However, the established assumptions about the statistical structure of
fMRI data that are built into standard univariate analyses do not give
reason to expect a gain from nonlinear algorithms. On the contrary, a
nonlinear classifier may confound different partial effects in a
factorial design that can be differentiated by the multivariate linear
approach proposed in this paper.

\begin{figure}[p]
\centering
\includegraphics{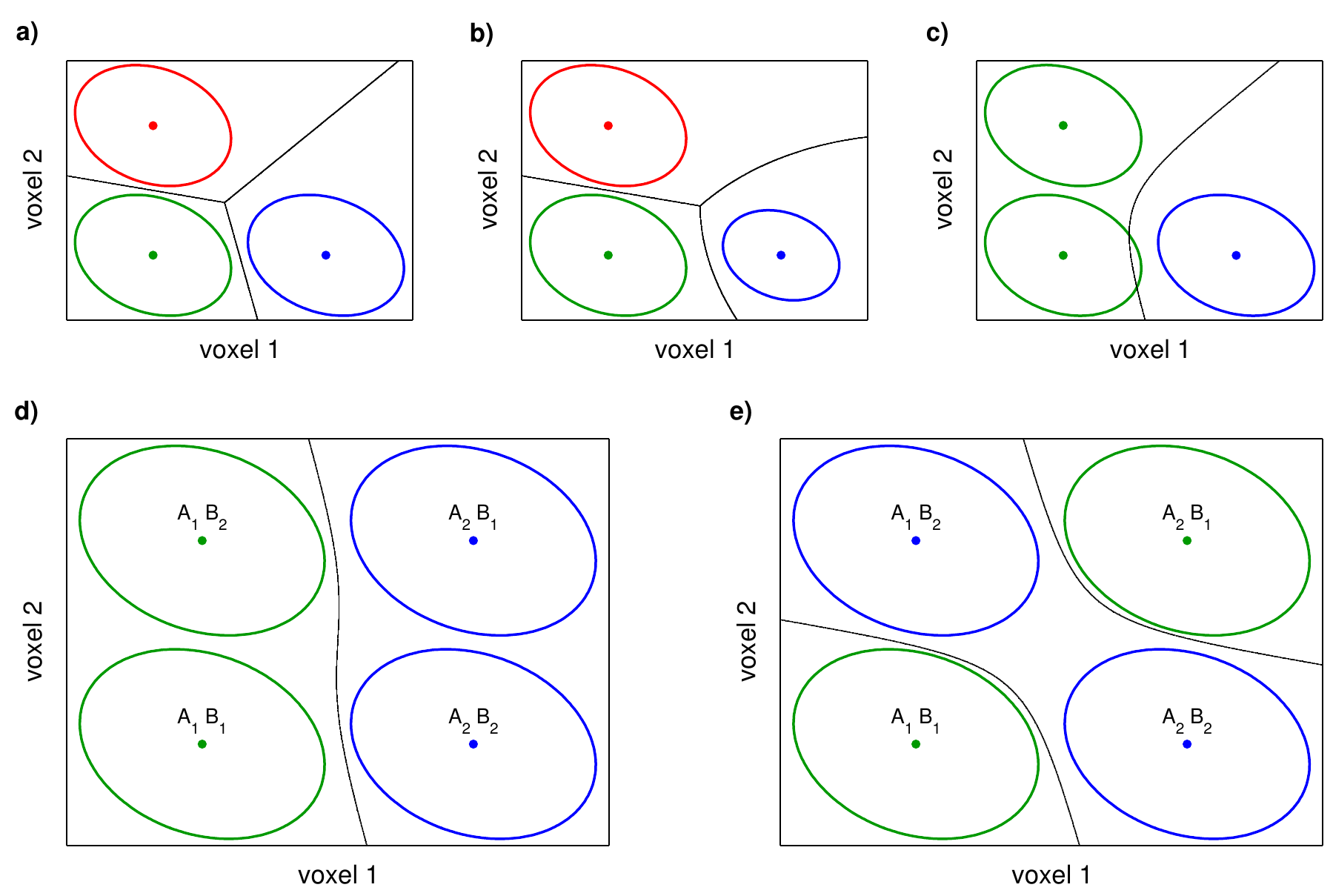}
\caption{\textbf{Causes for and misinterpretation of nonlinear optimal
classification.} \ \textbf{(a)}\ If the data adhere to the assumptions of
the multivariate linear model which implies that all classes are
characterized by the same elliptical distribution (red, green and blue
ellipses) with the same variance (homoscedasticity), optimal
classification boundaries (black lines) are also piecewise linear.
\ \textbf{(b)}\ These boundaries become nonlinear if the variance differs
between classes (violation of homoscedasticity), which however is
incompatible with the assumption of additive noise made by the MGLM
\eqref{mglm}. \ \textbf{(c)}\ A more realistic scenario for how
heteroscedasticity and thereby nonlinearity may arise is that one class
(green) is actually a compound of two or more patterns (green \emph{and}
red in panel a). \ \textbf{(d)}\ Such compound classes occur for instance
in the decoding of single factors in a multifactorial design. In the
case illustrated here there are two factors A and B with two levels
each. The four resulting conditions are pooled into two classes
according to the levels of the factor A (green for $\A_1$, blue for
$\A_2$). The clear separation of these classes by the classification
boundary is consistent with the fact that there is a multivariate main
effect of factor A. \ \textbf{(e)}\ However, a nonlinear classifier would
also be able to separate data pooled according to the levels of factor B
(green for $\B_1$, blue for $\B_2$), though there is no multivariate
main effect of factor B. By contrast, the information picked up by the
nonlinear classifier would be correctly identfied as a multivariate
interaction effect by the MANOVA.}
\end{figure}

For this it is important to note that nonlinear optimal classification
boundaries do not arise from a nonlinearity of the \emph{patterns}, but
from a different \emph{error} covariance structure within different
classes. While the pattern characterizing a class has presumably been
generated by a process of nonlinear neural dynamics (cf. Haken, 1995),
the pattern vector $\vec \mu$ itself is a point in voxel activation
space and therefore does not allow for nonlinear structure. If the data
points belonging to each class are distributed according to the same
elliptical distribution with different means (Fig.\ 7a), optimal
classification boundaries are piecewise linear. Nonlinear boundaries
(Fig.\ 7b) arise if this `homoscedasticity' assumption is violated. In
the simplest case, classification boundaries become (hyper-) paraboloid
(quadratic discriminant analysis; see Hastie et al., 2009).

A violation of homoscedasticity is not to be expected as long as the
within-class variance consists only of the contributions of complex
unknown background processes described by the additive error term $\Xi$
in the MGLM, and all systematically occurring effects are explicitly
modeled. The most likely cause of heteroscedasticity is therefore
insufficient modeling of the data. An example is shown in Fig.\ 7c: A
situation where there are actually three different patterns involved has
been modeled by two classes, disregarding the distinction between two of
the patterns. Because of this, the two resulting classes have a
different covariance structure, and one of them is also no longer
meaningfully characterized by a single pattern vector. Consequently, the
classification boundary becomes nonlinear. However, such a `pooling' of
data from different experimental conditions, which is sometimes used in
classifier-based MVPA studies utilizing more complex designs (e.g. Cichy
et al., 2012; Momennejad and Haynes, 2012), can be avoided in the MGLM
framework: The design matrix should always be constructed such that it
models all the systematic experimental effects; and the selection of
partial effects is implemented by choosing the correct contrast.

The classic nonlinear `XOR problem' illustrated in Fig.\ 7d\&e
demonstrates that the application of a nonlinear classifier to pooled
data can even be misleading. In this example, the four different
patterns belong to the four cells of a 2\ $\times$\ 2 factorial design,
and the two classes encode either the two levels of factor A (6d) or of
factor B (6e). The success of classification in the second case might be
interpreted such that there is an effect of factor B, since data pooled
according to the levels of this factor can be nonlinearly separated.
However, from a MANOVA perspective the situation shown is characterized
by an interaction A $\times$ B and a main effect of factor A, but no
main effect of factor B. The MGLM framework is therefore able to
describe the structure of the data in a more detailed way exactly
because it is linear.

\subsection{Insufficient data}\label{insufficient-data}

A limitation of MVPA measures like Mahalanobis distance or LDA
classification accuracy that use an explicit probabilistic model of the
fMRI data (a `generative' model) is that in a high-dimensional space
spanned by a large number of voxels the number of data points is not
sufficient to properly estimate the within-class covariance $\Sigma$. In
such a case, using a `discriminative' model, which does not describe the
distribution of the data points but only allows to assign them to
classes, may be necessary.

This limitation holds for the cross-validated MANOVA method presented in
this paper, too. However, the situation is vastly improved compared to
methods applied to run-wise parameter estimates because the MGLM
operates on the level of single volumes. Moreover, it makes even better
use of the data than trial-wise classification. Where a classifier would
assess the within-class variation based on the volumes belonging to the
two classes of trials involved, the error covariance of the MGLM is
estimated using the residuals of the full model across the whole length
of the recording. Since within the cross-validation scheme (Sec.\ 2.3)
the error covariance estimate in each fold is based on data from all but
one of the $m$ runs, a non-singular estimate $E$ is obtained as soon as
$(m - 1) \, f_E \geq p$, where the number of error degrees of freedom is
given by the number of volumes per run minus the number of linearly
independent regressors. For searchlight-based MVPA of typical fMRI
studies not just a non-singular but a good estimate should be possible
for radii up to $4$ ($p = 257$).

The estimation problem persists for the analysis of patterns in large
regions of interest or across the whole brain (considered already by
Friston et al., 1995). Here, regularization of the estimate using
`shrinkage' (see Blankertz et al., 2011; Schäfer and Strimmer, 2005) or
reduction of dimensionality via principal component analysis as a
preprocessing step (cf. Carlson et al., 2003) or via built-in model
constraints (Worsley et al., 1997) are possible solutions, but with the
drawback that the resulting estimator is biased. A better approach to
extend the field of application of cvMANOVA might therefore be to
develop a model of the local error covariance structure, e.g.\ based on
voxel distance or tissue types.

\section{Conclusion}\label{conclusion}

In this paper we have introduced a measure for use in searchlight-based
MVPA studies, the \emph{pattern distinctness} $D$, as a replacement of
the often employed measure of accuracy. Instead of quantifying the
performance of a classifier, our measure is based on the multivariate
extension of the general linear model (the MGLM). It therefore does not
depend on a particular classifier and its parameters, but directly
characterizes the structure of fMRI data by quantifying the amount of
multivariate variance explained by a given effect, in terms of the error
variance. The measure is related to standard MANOVA statistics, but
cross-validation is applied to obtain an unbiased estimate $\hat D$ of
the size of the effect. Other than the approach of Friston et al. (2008)
that implements Bayesian model selection for the MGLM, our method aims
to be a straightforward extension of univariate statistics to the
multivariate case.

The MGLM underlying cross-validated MANOVA uses the same design matrix
and is analyzed using the same contrast matrices as those used with the
GLM, which thereby provide a unified basis for univariate and
multivariate analyses. Because it is based on the MGLM, cvMANOVA can be
used for two or more classes of trials but also with parametric
regressors, and $\hat D$ can quantify the multivariate effect defined by
any estimable contrast. This way it becomes possible to use MVPA to
investigate main effects in a factorial design as well as interactions,
i.e.\ the question whether the \emph{pattern difference} between two or
more levels of one factor \emph{changes} depending on another factor. A
variant of the measure additionally allows to quantify \emph{pattern
stability} across the levels of a factor (corresponding to
`cross-decoding'). Our method therefore substantially increases the
number of studies where MVPA can be applied in a natural and simple way,
and the number of questions that can be asked of the data.

\section*{Appendix}

\appendix
\renewcommand{\theequation}{\Alph{section}.\arabic{equation}}
\setcounter{equation}{0}

\section{Estimation bias}\label{estimation-bias}

The expectation value of $\hat D_l$ is \[
\langle \hat D_l \rangle
= \tr \left ( \left\langle H_l \  E_l ^{-1} \right\rangle \right )
= \tr \left ( \left\langle H_l \right\rangle \  \langle E_l ^{-1} \rangle \right ),
\] because $H_l$ and $E_l$ are uncorrelated, since $\hat B_\Delta$ and
$\hat \Xi$ arise from mutually orthogonal projections.

For the hypothesis matrix, \[
\langle H_l \rangle
= \sum_{k \neq l} \left\langle \left \{ \hat B_\Delta' \right \}_k
\, \left \{ X' X \hat B_\Delta \right \}_l\right\rangle
= \sum_{k \neq l} \left\langle \left \{ \hat B_\Delta' \right \}_k \right\rangle
\left\langle \left \{ X' X \hat B_\Delta \right \}_l \right\rangle,
\] because the two braces are computed from data of different runs, and
since $\langle \hat B_\Delta \rangle = B_\Delta$, \[
\langle H_l \rangle = (m - 1) \, B_\Delta' X' X B_\Delta.
\]

The error matrix follows a Wishart distribution (Timm, 2002), \[
E_l = \sum_{k \neq l} \left \{ \hat \Xi' \hat \Xi \right \}_k
\quad \dist \Dist W_p \left ( \Sigma, (m - 1) \, f_E \right ),
\] and therefore the expectation of its inverse is
\[ \label{inversewishart}
\langle E_l^{-1} \rangle = \frac{1}{(m - 1) \, f_E - p - 1} \Sigma^{-1}.
\]

Consequently, \[
\langle \hat D_l \rangle
= \frac{(m - 1) \, n}{(m - 1) \, f_E - p - 1}
\  \tr \left ( \frac1n B_\Delta' X' X B_\Delta \Sigma^{-1} \right ),
\] where the trace term is identical to the definition of $D$
\eqref{trace}, and therefore \[
\langle \hat D \rangle
= \left\langle
\frac{(m - 1) \, f_E - p - 1}{(m - 1) \, n}
\cdot \frac1m \sum_{l = 1}^m \hat D_l
\right\rangle = D.
\]

\section{Null distribution}\label{null-distribution}

In order to calculate the variance of $\hat D_l$ for $B_\Delta = 0$, we
first approximate $E_l^{-1}$ by its expectation value
\eqref{inversewishart}, so that \[
\hat D_l \approx \frac{1}{(m - 1) \, f_E - p - 1}
\  \tr ( H_l \  \Sigma^{-1} )
\quad \text{for }
p \ll (m - 1) \, f_E.
\] For the trace holds \[
\tr ( H_l \  \Sigma^{-1} ) = \tr \left (
\sum_{k \neq l} \left \{ \hat B_\Delta' \right \}_k \ 
\left \{ X' X \hat B_\Delta \right \}_l
\cdot \Sigma^{-1} \right ),
\] and since \[
\hat B_\Delta = C C^- X^- \  Y = C C^- X^- \  \Xi
\] for $B_\Delta = 0$, \[
\tr ( H_l \  \Sigma^{-1} ) = \sum_{k \neq l}
\tr \left (
\left \{ \Xi' X'^- C C^- \right \}_k
\left \{ X' X C C^- X^- \Xi \right \}_l
\cdot \Sigma^{-1} \right ).
\] Under the trace, $\Sigma^{-1}$ acts as a normalization of $\Xi_k$ and
$\Xi_l$, so that we can choose $\Sigma = \I$ (the identity matrix)
without loss of generality.

Without the intermediate matrix expressions, $\tr ( \Xi_k' \Xi_l )$ is
the sum of $p$ inner products of two different $n$-dimensional random
vectors each, with elements independently distributed as
$\Dist N (0, 1)$. That is, it is the sum of $p \, n$ product-normally
distributed random variables. Due to the additional matrices forming
projection operators, the inner products are actually calculated in an
$f_H$-dimensional subspace (due to the contrast $C$) of a
$q$-dimensional space (due to the design matrix $X$), and the trace
becomes a sum of $p \, f_H$ independent product-normally distributed
random variables, each of unit variance. Therefore \[
\var \left ( \tr ( H_l \  \Sigma^{-1} ) \right )
= (m - 1) \  p \, f_H,
\] and \[
\var \hat D_l
\approx \frac{ (m - 1) \, p \, f_H }{ \left ((m - 1) \, f_E - p - 1 \right )^2 }.
\] Moreover, since due to the central limit theorem sums of independent
random variables tend towards the normal distribution, we can expect
$\hat D_l$ to be approximately normally distributed around 0 for larger
$p \, f_H$ (rule of thumb: $> 30$).

Due to complex statistical dependencies between the cross-validation
folds the simple extension,
$\var \hat D \approx p \, f_H / ((m - 1) \, m \, n^2)$, can only be a
very rough approximation. However, the proportionality to $p$ should
hold well for $p \ll (m - 1) \, f_E$.

\section{Pattern stability}\label{pattern-stability-1}

We are interested in the stability of an effect $\E$ encoded by a
contrast matrix $C$, across the $L$ levels of a factor $\A$. $C$ may be
an arbitrary contrast involving one or more other factors of a factorial
design, but not $\A$. It therefore consists of a `contrast element' $c$
that is replicated across the levels of $A$ \[
C = \left ( \begin{array}{c}
c \\ c \\ \vdots \\ c
\end{array} \right ).
\] We denote as $C_i$ the partial contrasts of $C$, i.e versions of $C$
where all the replications of $c$ are replaced by zeros, except for the
one associated with level $i = 1 \ldots L$ of factor $\A$.

The matrix to extract the parameter difference $B_\Delta$ associated
with effect $\E$ from the parameters $B$ of the full model
\eqref{extract} has the form of a Kronecker product \[
\mathcal C_\E = C C^- = \frac1L
\left ( \begin{array}{cccc}
1 & 1 & \ldots & 1 \\
1 & 1 & \ldots & 1 \\
\vdots & \vdots & \ddots & \vdots \\
1 & 1 & \ldots & 1
\end{array} \right )
\  \otimes \  (c c^-),
\] while the corresponding matrix for the interaction $\E \times \A$ has
the form \[
\mathcal C_{\E \times \A} =  \frac1L
\left ( \begin{array}{cccc}
(L - 1) & -1 & \ldots & -1 \\
-1 & (L - 1) & \ldots & -1 \\
\vdots & \vdots & \ddots & \vdots \\
-1 & -1 & \ldots & (L - 1)
\end{array} \right )
\  \otimes \  (c c^-).
\]

The condition of maximal inconsistency of the multivariate effects
associated with $c$ across the levels of $\A$ is defined by the mutual
orthogonality of the partial effects, $C_i C_i^- \  B$. In this case,
contributions due to the off-diagonal elements of the matrices in the
previous two expressions become zero, and consequently \[
D(\E) = \frac1{L - 1} \  D(\E \times \A).
\]

The resulting measure of pattern stability, \[
D(\E \setminus \A) = D(\E) - \frac1{L - 1} \  D(\E \times \A),
\] can be considered a special form of the pattern distinctness $D$
which is defined by an extraction matrix \[
\mathcal C_{\E \setminus \A} = \mathcal C_\E - \frac1{L - 1} \  \mathcal C_{\E \times \A} =  \frac1{L - 1}
\left ( \begin{array}{cccc}
0 & 1 & \ldots & 1 \\
1 & 0 & \ldots & 1 \\
\vdots & \vdots & \ddots & \vdots \\
1 & 1 & \ldots & 0
\end{array} \right )
\  \otimes \  (c c^-).
\] In this form the analogy to cross-decoding becomes apparent: The
measure only comprises terms that combine patterns from \emph{different
levels} of the factor $\A$. $\mathcal C_{\E \setminus \A}$ cannot be
written as a product $C C^-$ and therefore does not correspond to a
contrast, but $D(\E \setminus \A)$ shares the statistical properties of
the pattern distinctness $D$.

\section{Permutation statistics}\label{permutation-statistics}

As shown in Appendix A, if there is no true effect the single-fold
estimate $\hat D_l$ is asymptotically normally distributed around $0$.
However, the approximation we were able to derive for $\var \hat D_l$
does not simply translate to $\hat D$ itself, and we do not have any
results for higher order moments. Combined with the fact that empirical
data will not exactly adhere to the assumption of multivariate normality
underlying these derivations, we recommend to use a permutation
procedure for the assessment of a statistically significant deviation
from a null hypothesis, $\Hnull: D = 0$.

A permutation test (Good, 2005; Lehmann and Romano, 2005; Nichols and
Holmes, 2002) is a null hypothesis test that makes only weak
distributional assumptions, and derives the critical value of a test
statistic by a computational procedure that utilizes the given sample.
The method is based on the symmetries of the distribution of the data
(or derived statistics) under $\Hnull$, which make it possible to
generate a set of artificial samples consistent with the null hypothesis
by exchanging (`permuting') parts of the original data. The test
statistic is computed for all possible permutations, and the null
hypothesis is rejected if the actual value of the statistic has an
extreme position within the set of permutation values.

In the case of cross-validated MANOVA, we propose to implement
permutations at the level of experimental runs beause they can be
considered mutually statistically independent. The circumstance that
MGLM parameters $B$ are estimated separately for each run and the
cross-validation procedure combines parameter estimates from different
runs into the measure $\hat D$ leads to the following approach:

Under the null hypothesis $D = 0$ for a given contrast $C$, the contrast
parameter estimate $\hat B_\Delta = C C^- \hat B$ is symmetrically
distributed around $0$. The resulting \emph{sign permutation} \[
\left \{ \hat B_\Delta \right \}_k
\quad \rightarrow \quad
- \left \{ \hat B_\Delta \right \}_k
\] applies independently for each run $k = 1 \ldots m$, so that formally
there are $2 ^ m$ permutations. Since reversing signs for all runs at
once does not change the value of $\hat D$, half of these permutations
are redundant, leaving an effective number of $2 ^ {m - 1}$
permutations.

For the typical number of runs of an fMRI experiment, the resulting
number of permutations is not sufficient to perform a null hypothesis
test for a single subject at standard significance levels. In this
paper, we follow the approach of Stelzer et al. (2013) to construct a
group-level permutation distribution by combining permutations
independently selected in each subject (see Sec.\ 3).

\section*{Acknowledgements}

The authors would like to thank Guillaume Flandin for providing
invaluable insights into the innards of SPM8 and Radek Cichy for access
to the fMRI data set, Martin Hebart, Kerstin Hackmack, María Herrojo
Ruiz, Carsten Bogler, Martin Weygand, Jakob Heinzle, Stefan Bode, Kai
Görgen, and Fernando Ramírez for discussions, comments, and help, as
well as two anonymous reviewers for helpful criticism.

An implementation of the analysis method for use with SPM8 can be
obtained from the corresponding author.

\section*{References}

Abrams, D., Bhatara, A., Ryali, S., Balaban, E., Levitin, D., Menon, V.,
2011. Decoding temporal structure in music and speech relies on shared
brain resources but elicits different fine-scale spatial patterns.
Cereb. Cortex 21, 1507--1518.

Blankertz, B., Lemm, S., Treder, M., Haufe, S., Müller, K.-R., 2011.
Single-trial analysis and classification of eRP components -- a
tutorial. NeuroImage 56, 814--825.

Boser, B.E., Guyon, I.M., Vapnik, V.N., 1992. A training algorithm for
optimal margin classifiers, in: Proceedings of the Fifth Annual Workshop
on Computational Learning Theory. ACM, pp. 144--152.

Carlson, T.A., Schrater, P., He, S., 2003. Patterns of activity in the
categorical representations of objects. J. Cogn. Neurosci. 15, 704--717.

Chang, C.-C., Lin, C.-J., 2011. LIBSVM: a library for support vector
machines. ACM Trans. Intell. Syst. Technol. 2, 27:1--27:27.

Cichy, R.M., Chen, Y., Haynes, J.D., 2011. Encoding the identity and
location of objects in human LOC. NeuroImage 54, 2297--2307.

Cichy, R.M., Heinzle, J., Haynes, J.D., 2012. Imagery and perception
share cortical representations of content and location. Cereb. Cortex
22, 372--380.

Cohen, J., 1982. Set correlation as a general multivariate data-analytic
method. Multivariate Behavioral Research 17, 301--341.

Cohen, J., 1988. Statistical power analysis for the behavioral sciences,
2nd ed. Lawrence Erlbaum.

Cortes, C., Vapnik, V., 1995. Support-vector network. Mach. Learn. 20,
273--297.

Cox, D.D., Savoy, R.L., 2003. Functional magnetic resonance imaging
(fMRI) ``brain reading'': detecting and classifying distributed patterns
of fMRI activity in human visual cortex. NeuroImage 19, 261--270.

Edelman, S., Grill-Spector, K., Kushnir, T., Malach, R., 1998. Toward
direct visualization of the internal shape representation space by fMRI.
Psychobiology 26, 309--321.

Fang, K., Kotz, S., Ng, K., 1990. Symmetric multivariate and related
distributions, ed. Chapman; Hall.

Friston, F., Holmes, A.P., Worsley, K.J., Poline, J.-P., Frith, C. D.,
Frackowiak, R.S.J., 1995. Statistical parametric maps in functional
imaging: a general linear approach. Hum. Brain Mapp. 2, 189--210.

Friston, K., Chu, C., Mourão-Miranda, J., Hulme, O., Rees, G., Penny,
W., Ashburner, J., 2008. Bayesian decoding of brain images. NeuroImage
39, 181--205.

Friston, K.J., Frith, C.D., Liddle, P.F., Frackowiak, R.S., 1993.
Functional connectivity: the principal-component analysis of large (PET)
data sets. J. Cereb. Blood Flow Metab. 13, 5--14.

Glaser, D., Friston, K., 2007. Covariance components, in: Friston, K.J.,
others (Eds.), Statistical Parametric Mapping: the Analysis of
Functional Brain Images. Academic Press.

Good, P.I., 2005. Permutation, parametric, and bootstrap tests of
hypotheses, 3rd ed. Springer.

Goutte, C., Toft, P., Rostrup, E., Nielsen, F.Å., Hansen, L.K., 1999. On
clustering fMRI time series. NeuroImage 9, 298--310.

Haken, H., 1995. Principles of brain functioning: a synergetic approach
to brain activity, behavior, and cognition, ed. Springer.

Hastie, T., Tibshirani, R., Friedman, J., 2009. The elements of
statistical learning: data mining, inference and prediction, 2nd ed.
Springer.

Haxby, J.V., 2012. Multivariate pattern analysis of fMRI: the early
beginnings. NeuroImage.

Haxby, J.V., Gobbini, M.I., Furey, M.L., Ishai, A., Schouten, J.L.,
Pietrini, P., 2001. Distributed and overlapping representations of faces
and objects in ventral temporal cortex. Science 293, 2425--2430.

Haynes, J.-D., Rees, G., 2005a. Predicting the orientation of invisible
stimuli from activity in human primary visual cortex. Nat. Neurosci. 8,
686--691.

Haynes, J.-D., Rees, G., 2005b. Predicting the stream of consciousness
from activity in human visual cortex. Curr. Biol. 15, 1301--1307.

Haynes, J.-D., Rees, G., 2006. Decoding mental states from brain
activity in humans. Nat. Rev. Neurosci. 7, 523--534.

Kahnt, T., Grueschow, M., Speck, O., Haynes, J.-D., 2011. Perceptual
learning and decision-making in human medial frontal cortex. Neuron 70,
549--559.

Kamitani, Y., Tong, F., 2005. Decoding the visual and subjective
contents of the human brain. Nat. Neurosci. 8, 679--685.

Kiebel, S.J., Holmes, A.P., 2007. The general linear model, in: Friston,
K.J., others (Eds.), Statistical Parametric Mapping: the Analysis of
Functional Brain Images. Academic Press.

Kriegeskorte, N., Goebel, R., Bandettini, P., 2006. Information-based
functional brain mapping. Proc. Natl. Acad. Sci. U. S. A. 103,
3863--3868.

Kruggel, F., Cramon, D.Y. von, 1999. Modeling the hemodynamic response
in single-trial functional mRI experiments. Magn. Reson. Med. 42,
787--797.

Kullback, S., Leibler, R.A., 1951. On information and sufficiency. Ann.
Math. Stat. 22, 79--86.

Lehmann, E.L., Romano, J.P., 2005. Testing statistical hypotheses, 3rd
ed. Springer.

Mahalanobis, P.C., 1936. On the generalised distance in statistics.
Proc. Natl. Inst. Sci. India 2, 49--55.

McIntosh, A.R., Bookstein, F.L., Haxby, J.V., Grady, C.L., 1996. Spatial
pattern analysis of functional brain images using partial least squares.
NeuroImage 3, 143--157.

McKeown, M.J., Makeig, S., Brown, G.G., Jung, T.-P., Kindermann, S.S.,
Bell, A.J., Sejnowksi, T.J., 1998. Analysis of fMRI data by blind
separation into independent spatial components. Hum. Brain Mapp. 6,
160--188.

Mitchell, T.M., Hutchinson, R., Niculescu, R.S., Pereira, F., Wang, X.,
Just, M., Newman, S., 2004. Learning to decode cognitive states from
brain images. Mach. Learn. 57, 145--175.

Momennejad, I., Haynes, J.-D., 2012. Human anterior prefrontal cortex
encodes the ``what'' and ``when'' of future intentions. NeuroImage 61,
139--148.

Nichols, T.E., Holmes, A.P., 2002. Nonparametric permutation tests for
functional neuroimaging: a primer with examples. Hum. Brain Mapp. 15,
1--25.

Norman, K.A., Polyn, S.M., Detre, G.J., Haxby, J.V., 2006. Beyond
mind-reading: multi-voxel pattern analysis of fMRI data. Trends Cognit.
Sci. 10, 424--430.

Pereira, F., Mitchell, T., Botvinick, M., 2009. Machine learning
classifiers and fMRI: a tutorial overview. NeuroImage 45, S199--S209.

Schäfer, J., Strimmer, K., 2005. A shrinkage approach to large-scale
covariance matrix estimation and implications for functional genomics.
Stat. Appl. Genet. Mol. Biol. 4, 32.

Stelzer, J., Chen, Y., Turner, R., 2013. Statistical inference and
multiple testing correction in classification-based multi-voxel pattern
analysis (mVPA): random permutations and cluster size control.
NeuroImage 65, 69--82.

Timm, N.H., 2002. Applied multivariate analysis, ed. Springer.

Tong, F., Pratte, M.S., 2012. Decoding patterns of human brain activity.
Annu. Rev. Psychol. 63, 483--509.

Wolpaw, J., Wolpaw, E.W. (Eds.), 2012. Brain--computer interfaces:
principles and practice, ed. Oxford University Press.

Worsley, K.J., Poline, J.-B., Friston, K.J., Evans, A.C., 1997.
Characterizing the response of pET and fMRI data using multivariate
linear models. NeuroImage 6, 305--319.

Wyman, F.J., Young, D.M., Turner, D.W., 1990. A comparison of asymptotic
error rate expansions for the sample linear discriminant function.
Pattern Recognit. 23, 775--783.

\end{document}